# Effects of Coronal Mass Ejections on Distant Coronal Streamers

B. Filippov[1], P. Kayshap[2], A.K. Srivastava[3], and O. Martsenyuk[1]

[1]*Pushkov Institute of Terrestrial Magnetism, Ionosphere and Radio Wave Propagation, Russian Academy of Sciences, Troitsk, Moscow, 142190 Russia*
*(e-mail:* bfilip@izmiran.ru*)*
[2]*Aryabhatta Research Institute of Observational Sciences (ARIES), Manora Peak, Nainital-263 129, Uttarakhand, India*
[3]*Department of Physics, Indian Institute of Technology (Banaras Hindu University), Varanasi, 221005, India*
*(e-mail:* aks@aries.res.in*)*

**Abstract**  The effects of a large coronal mass ejection (CME) on a solar coronal streamer located roughly 90° from the main direction of the CME propagation observed on January 2, 2012 by the SOHO/LASCO coronagraph are analyzed. Radial coronal streamers undergo some bending when CMEs pass through the corona, even at large angular distances from the streamers. The phenomenon resembles a bending wave traveling along the streamer. Some researchers interpret these phenomena as the effects of traveling shocks generated by rapid CMEs, while others suggest they are waves excited inside the streamers by external impacts. The analysis presented here did not find convincing arguments in favor of either of these interpretations. It is concluded that the streamer behavior results from the effect of the magnetic field of a moving magnetic rope associated with the coronal ejection. The motion of the large-scale magnetic rope away from the Sun changes the surrounding magnetic field lines in the corona, and these changes resemble the half-period of a wave running along the streamer.

**Keywords:** Coronal mass ejections (CMEs); Coronal rays; Streamers; Magnetic fields; Shock waves; Kink waves

## 1. Introduction

Coronal streamers feel an influence of coronal mass ejections (CMEs) propagating through the corona rather far from the position of the streamers. They are deflected and become temporarily curved (Hundhausen, et al., 1987; Sheeley, Hakala, and Wang, 2000; Tripathi and Raouafi, 2007; Vourlidas and Ontiveros, 2009; Filippov and Srivastava, 2010). The hump, created on the previously straight axis of the streamer, moves along it outwards as a kink perturbation. Many authors interpret this pattern as a manifestation of a shock produced by fast, super-Alfvenic CME propagation in the corona (e.g., Hundhausen, 1987; Sime and Hundhausen, 1987; Sheeley, Hakala, and Wang, 2000; van der Holst, van Driel-Gesztelyi, and Poedts, 2002; Tripathi and Raouafi, 2007; Vourlidas and Ontiveros, 2009). Deceleration should be a typical feature of the shock-produced kink, because the shock wave loses energy and slows down as it propagates through the corona.

Chen et al. (2010), Feng et al. (2011) consider the hump propagation along the streamer stalk as a streamer wave, the fast kink body mode carried by and propagating outwards along the streamer plasma sheet structure in the wake of the CME-caused streamer deflections. They found several candidates for streamer wave events in the Large Angle and Spectrometric Coronagraph (LASCO) (Brueckner et al. 1995) observations aboard the Solar and Heliospheric Observatory (SOHO) in Solar Cycle 23. All of them were associated with wide and fast CMEs with a linear speed no less than ∼1000 km s$^{-1}$.

In this paper, we analyze in detail the interaction of a CME with the remote coronal ray located nearly 90º away from the CME as observed by the SOHO/LASCO coronagraph on 2 January 2012 in order to find indications of shocks or streamer waves. We compare the results with our model of a moving flux rope associated with the CME.

## 2. Event of 2 January 2012

The fast wide CME appeared in the field of view of the LASCO C2 coronagraph at 15:12 UT on 2 January 2012 (Figure 1). It moved in the West direction at a speed of about 1100 km s$^{-1}$ (the position angle $P$ was about 270°). According to *SOHO*/LASCO CME Catalog[1], the CME was of halo type and showed a small deceleration of ~ 8 m s$^{-2}$. Although the angular width of the CME in the Catalog is shown as 360°, obviously (see Figure1 and Figure 2) its brightest part spreads in western direction. In the field-of-view of the LASCO C2 (2.5 $R \div 6\ R$), the leading edge of the CME moving in SWW direction accelerates with the acceleration of about 180 m s$^{-2}$ from 840 km s$^{-1}$ to 1100 km s$^{-1}$.

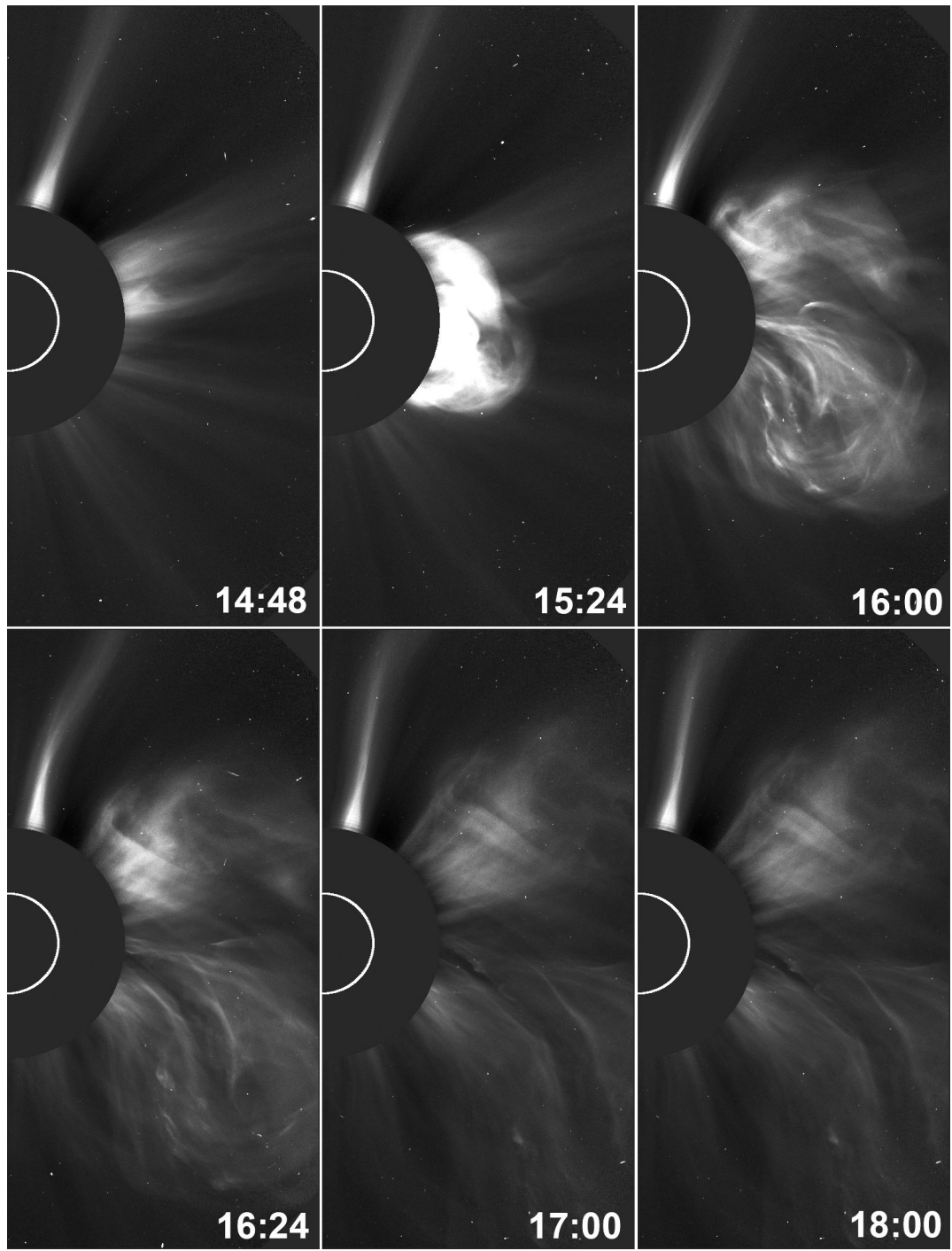

---

[1] http://cdaw.gsfc.nasa.gov/CME_list/

*Figure 1.* Northern coronal streamer deflection by the CME on 02 January 2012 in SOHO/LASCO C2 images (Courtesy:- SOHO/LASCO).

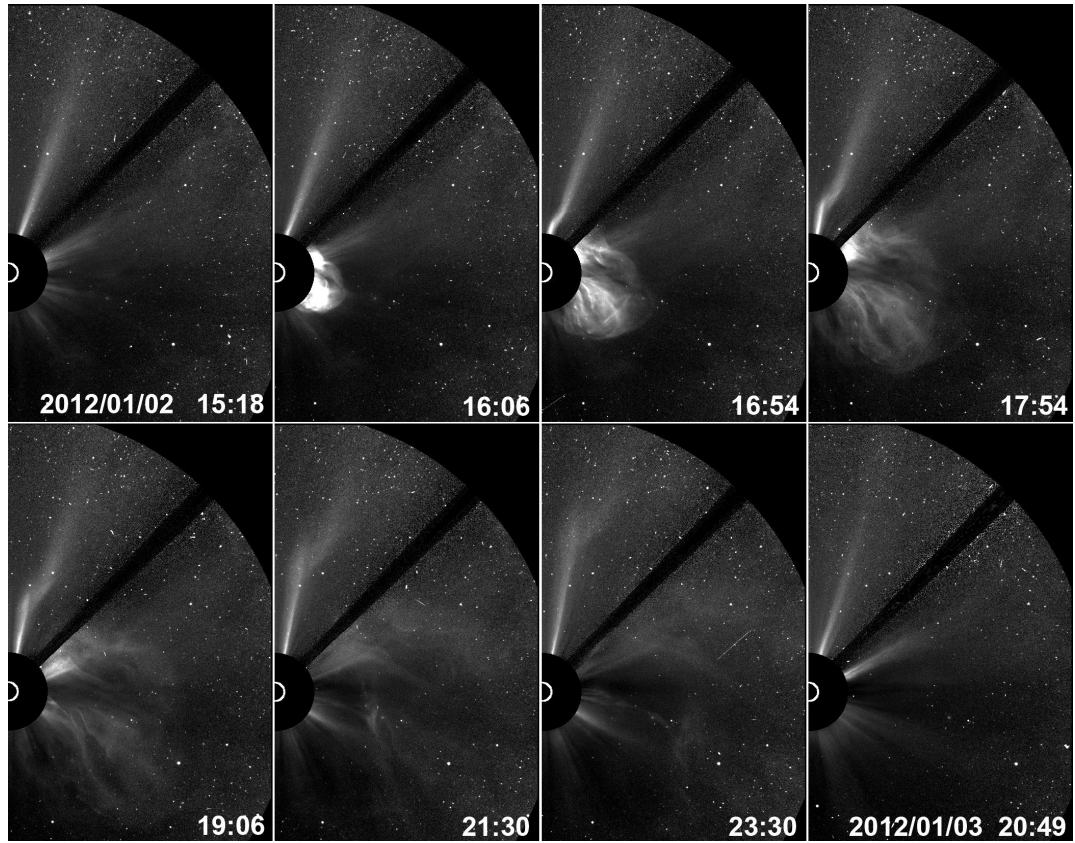

*Figure 2.* Northern coronal streamer deflection by the CME on 02 January 2012 in SOHO/LASCO C3 images (Courtesy:- SOHO/LASCO).

The streamer that was deflected by the CME, was located at $P = 342°$. In the field-of-view of the LASCO C2 (Figure 1), the streamer axis curved smoothly after 15:36 UT. The gentle hump moved along the ray and leaved the LASCO C2 field-of-view at 17:24 UT. In the field-of-view of the LASCO C3 (Figure 2), the event looks like a new streamer grows at new position closer to the North pole along with the propagation of the CME. The "old streamer" disappears at radial distances where the "new streamer" exists, while it is unchanging at greater distances. Between these two parts of the streamer, there is a slanted connecting section, which moves along with the growth of the new streamer. The position angle of the "new streamer" is $P = 346°$. It slowly returns to the former position $P = 342°$ (compare frames at 15:18 UT on 2 January and at 20:49 UT on 3 January in Figure 2).

For detailed analysis of streamer displacements, we construct maps from the circular strips of coronal images, which are concentric to the solar disk. We cut the strips through narrow circular slits (Figure 3) at different radial distances. The strips are then straightened and placed side by side in chronological order. The slits cover the position angles from 270° to 360° or latitudes from 0° to 90°. At the slice-time diagram (Figure 4) each strip corresponds to 12-minute interval in accordance with the cadence of LASCO C2 observations.

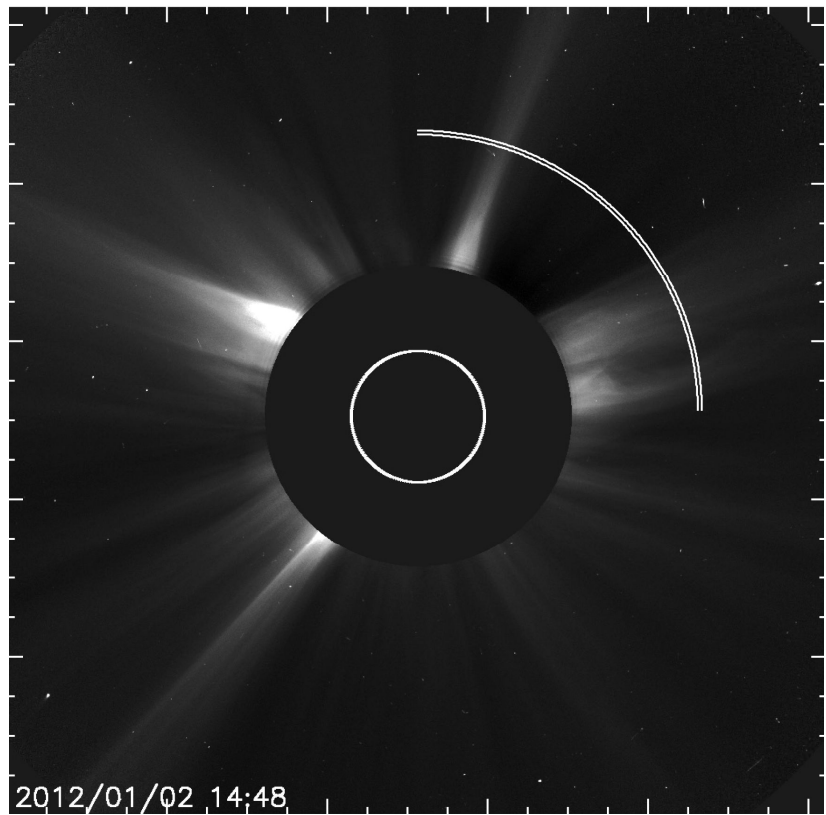

*Figure 3.* Position of a circular slit at the radial distance of 4.2 *R* in the LASCO C2 coronal image (Courtesy:- SOHO/LASCO).

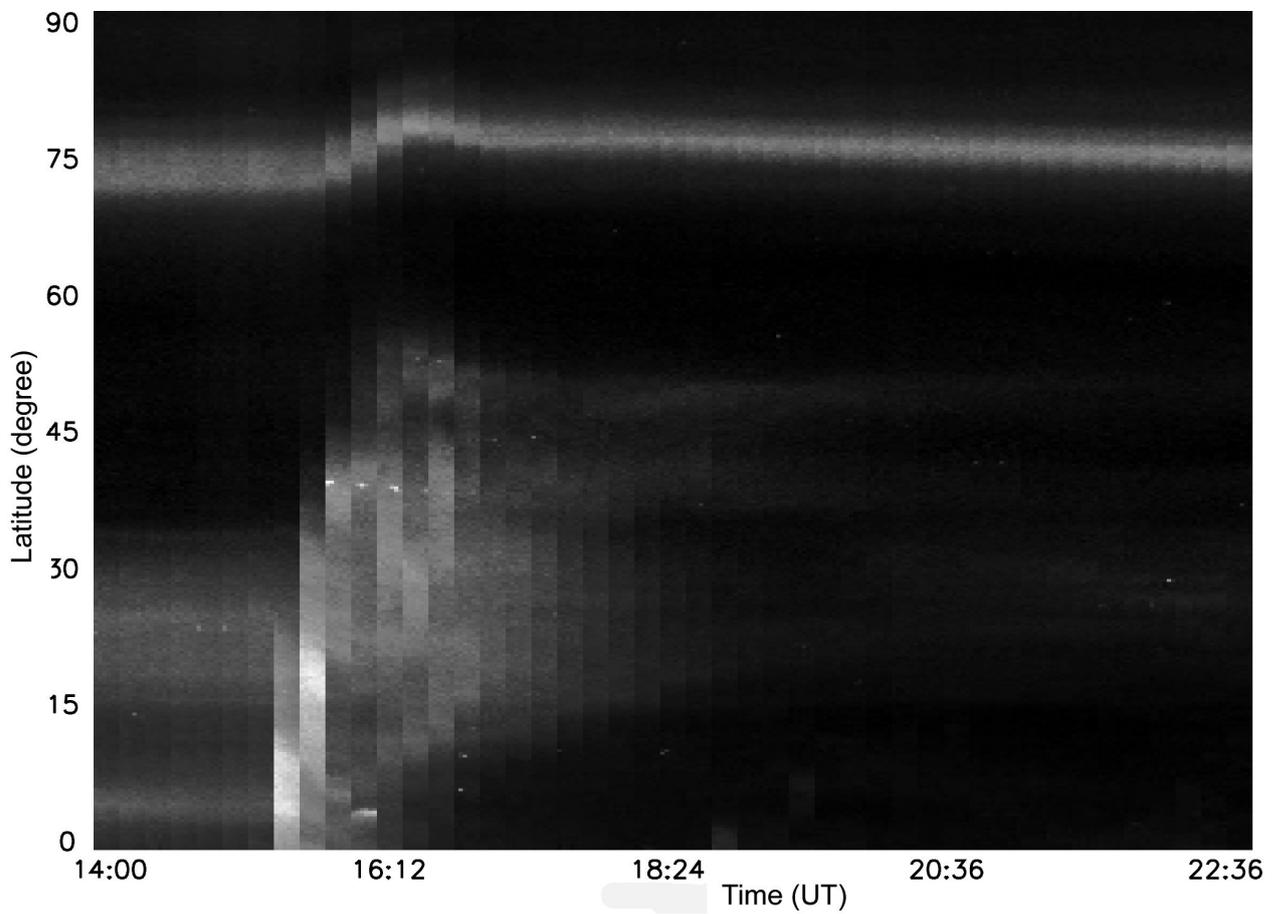

*Figure 4.* Slice-time map for the slit position at the radial distance of 3.3 *R*.

In Figure 4, we see the appearance of the bright CME at low latitudes, which spreads later on to higher latitudes. The streamer "feels" the presence of the CME after 24 minutes. Its axis moves to the North during 48 minutes, then goes back at first nearly at the same rate and afterwards very gradually to the initial position. No traces of streamer axis oscillations after the first impact are seen in Figure 4.

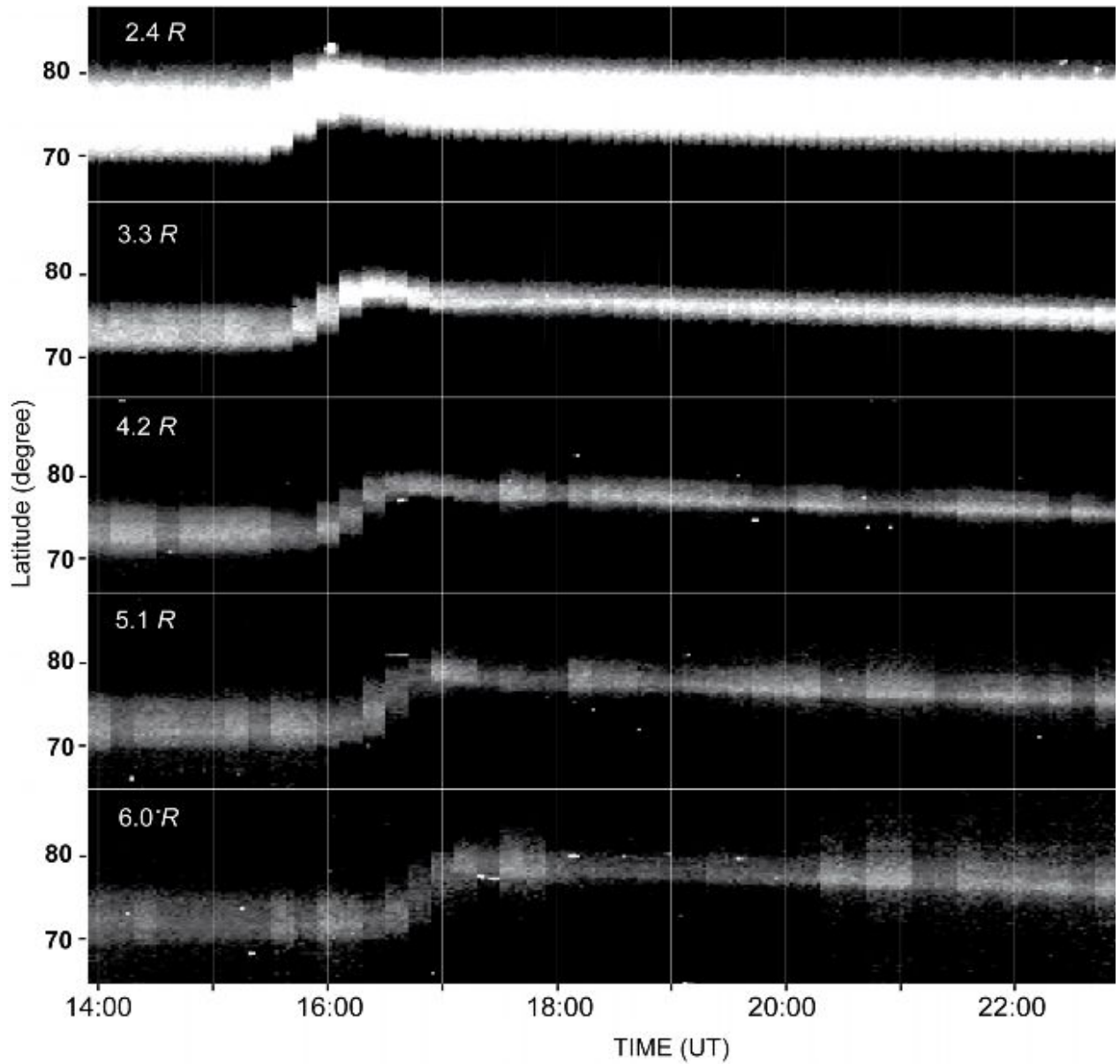

*Figure 5.* Time profiles of the deflection of the streamer at the heliocentric distances 2.4 *R*, 3.3 *R*, 4.2 *R*, 5.1 *R*, and 6.0 *R*.

Figure 5 shows the time profiles of the streamer bright core deflection at five heliocentric distances. The initial angular position slightly varies at different heights because of non-symmetric brightness distribution along a cross-section of the rather wide streamer. It is not related to possible non-radial direction or curvature of the streamer. The bump on the tops of the deflection curves becomes less prominent at greater distances. The northern edge of the CME also shows such a bump (see Figure 4), which also becomes significantly more smoothed at greater distances. In running differential images (Figure 6), the top of the bright ray, which corresponds to the deflected part of the streamer, moves synchronously with the leading edge of the northern part of the CME. The height-time plot of the bright ray top fairly well coincides with the time profile of the CME leading edge at $P = 310°$ (Figure 7). The speed of the CME leading edge in this direction increases from 550 km s$^{-1}$ to 850 km s$^{-1}$ in the field-of-view of the LASCO C2, while the bright ray slightly accelerates from 650 km s$^{-1}$ to 900 km s$^{-1}$.between 2.5 $R$ and 6 $R$. From the time difference between the profiles in Figure 5 one can derive that the beginning of the deflection propagates along the streamer with a speed of around 850 km s$^{-1}$ between 2.4 $R$ and 5.1 $R$ and around 450 km s$^{-1}$ between 5.1 $R$ and 6.0 $R$. The top of the kink moves slower with a speed around 500 km s$^{-1}$ because the shape of the kink becomes flatter with time.

The height-time map in Figure 8 is constructed by stacking narrow radial strips of LASCO C3 images at the position angle $P = 349°$ side by side. It shows the growth of the leading edge of the deflected streamer as a function of time. The border between dark and light areas has slightly downward curvature, indicating deceleration (from 700 km s$^{-1}$ to 450 km s$^{-1}$). The slope is greater in the lower corona. All aforecited velocity measurements are rather uncertain for several reasons. Using different techniques and tracking different points of a kink profile, we obtain unequal values. Different segments of the CME frontal structure move with different speeds and accelerations. We should keep this in mind when compare the speed of particular features in the event.

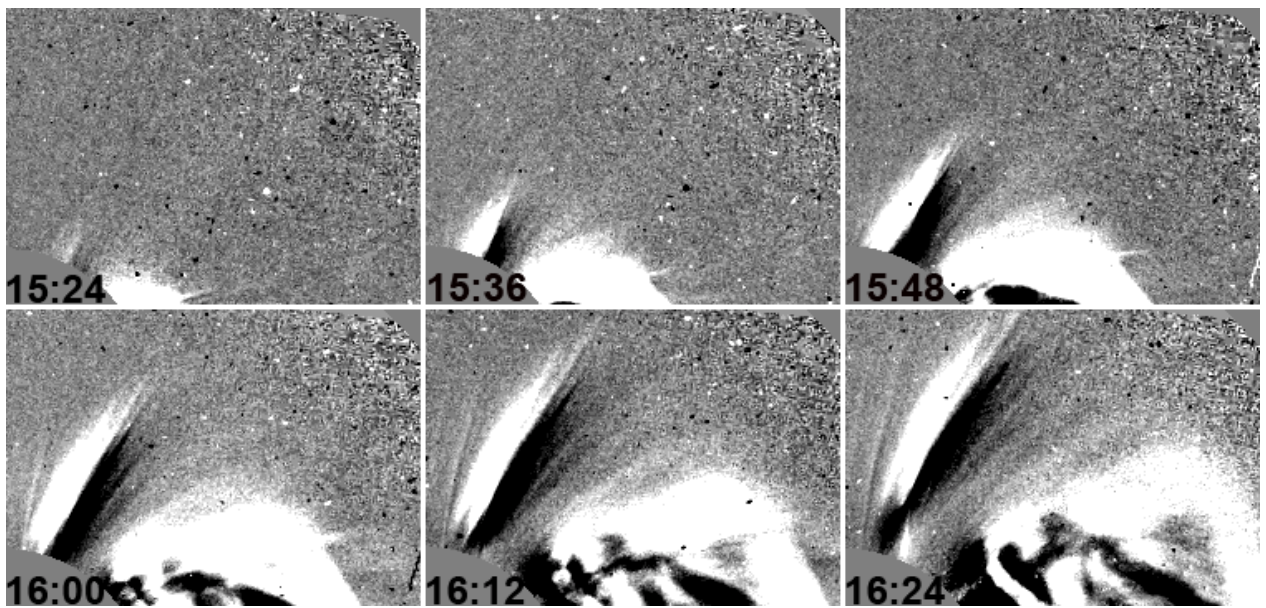

*Figure 6.* Running differential images of the streamer and the northern part of the CME.

The streamer deflection amplitude increases with a radial distance even in an angular measure as seen in Figure 5. The linear deflection at 6.0 $R$ is five times greater than that at 2.4 $R$. This tendency was observed also in the events studied by Filippov and Srivastava (2010).

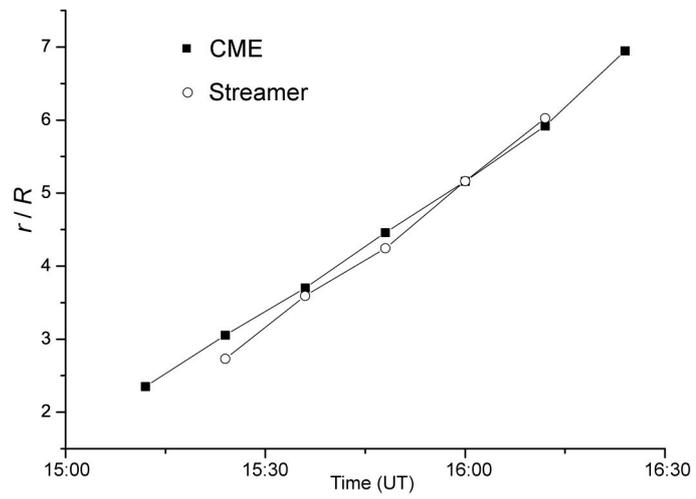

*Figure 7.* Height-time plot of the CME leading edge at $P = 310°$ and of the top of the curved streamer.

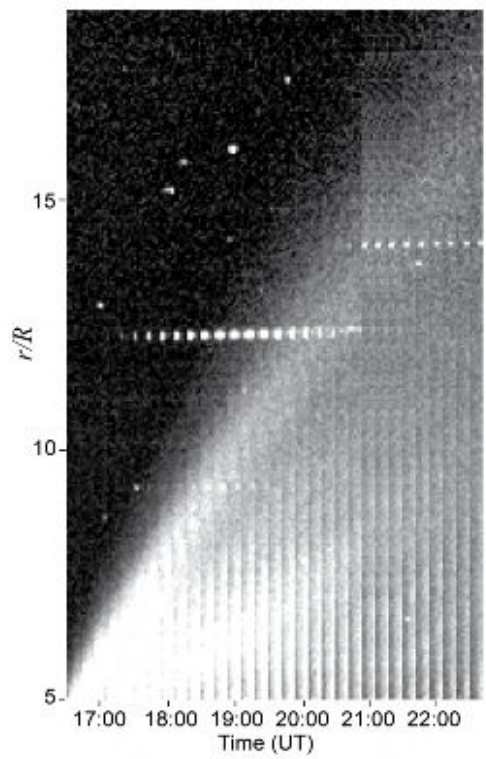

*Figure 8.* Height-time map of the leading edge of the "new streamer".

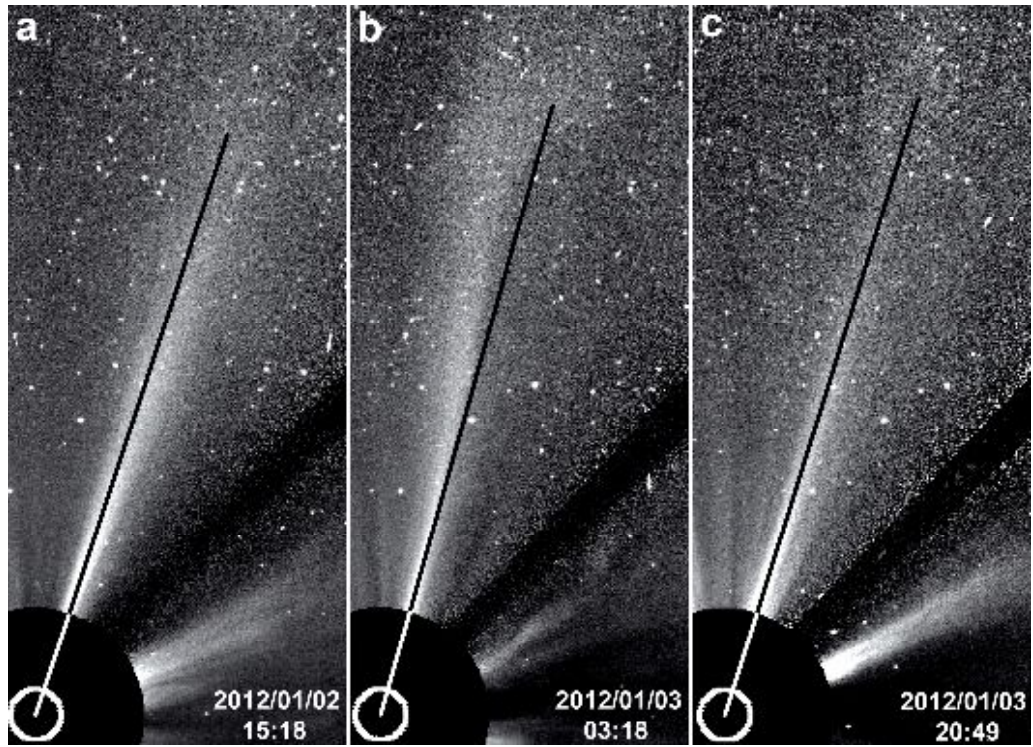

*Figure 9.* Streamer position relative to the radial direction at different times in the field-of-view of the LASCO C3 (Courtesy:- SOHO/LASCO).

The streamer before the CME and several hours after it shows rather perfect radial shape (Figure 9. But just after the passage of the CME when the major disturbance seems to have leaved the coronagraph field-of-view, the streamer deviates from the radial orientation (Figure 9(b)). This is confirmed also by the time profiles in Figure 5. At closer to the Sun distances, the streamer at first is restoring faster than at greater distances. That is why we have more bulging shape of the time profiles at lower heights and the curved shape of the streamer axis.

### 3. Discussion

How can we interpret the observed behavior of the streamer shape basing on the previous studies? First of all we should firmly state that there is no any evidence of streamer axis oscillations as in the events described by Chen *et al.* (2010) and Feng et al. (2011). The streamer axis deviates impulsively, then goes back at first nearly at the same rate in the lower section and afterwards returns very gradually to the initial position.

The CME that causes the streamer deflection is fast and very likely super-Alfvenic. No doubt, it is able to produce a shock wave in the corona. The shock might be responsible for the streamer deflection as supposed Hundhausen (1987). Sheeley, Hakala, and Wang (2000) suggested that the shock wave loses energy and slows down as it pushes obliquely across the radial magnetic field. That is why the kink produced by a remote fast CME would show deceleration. At the first glance, this seems to be consistent with the observations of our event. However, several features are not in line with the shock wave nature of the kink. This is the gradual return of the streamer to the pre-eruption position for a period of several hours and increase of the amplitude of the streamer displacement with radial distance. The letter was also observed in all events studied by Filippov and Srivastava (2010). It is difficult to perceive that the influence of the shock wave is so long and its action on the streamer is increasing at greater distances despite the losing energy by the shock and slowing down.

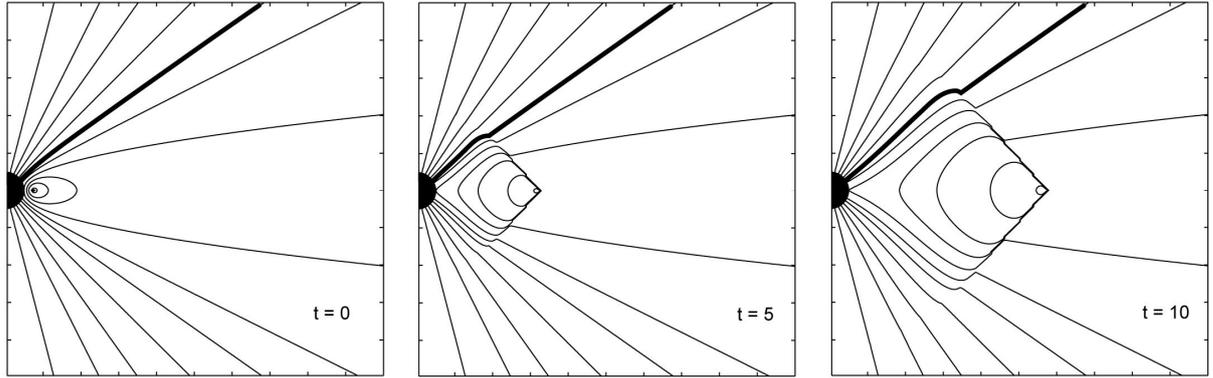

*Figure 10.* Field lines of a ring electric current moving with the velocity $v_c = 1.5\, v_a$ in the radial magnetic field.

We might interpret the observed streamer deflection as the influence of the magnetic field of a moving flux rope associated with a CME as suggested by Filippov and Srivastava (2010). Hundhausen (1987) concluded long ago that most of streamer deflections are caused by compressive magnetoacoustic waves moving out from the sides of CMEs approximately transverse to the nearly radial magnetic field. Sheeley, Hakala, and Wang (2000) pointed out that in the case of slow, accelerating CMEs that push aside streamers, the deflections are more permanent and they do not decelerate but instead share the accelerating CME motions. Figure 10 shows field lines of a ring electric current moving with the velocity $v_c$ higher than the Alfvén speed $v_a$ in the radial magnetic field (Filippov and Srivastava, 2010). The bold line can be considered as the streamer axis. Its shape is similar to the streamer curvature shown in Figure 9 (b). However, we cannot directly apply this simple model to our event. The nose of the CME moves in SWW direction, while the streamer nearly perpendicular to this direction. If the coronal current is concentrated within thin toroidal volume of the flux rope (ring current), its magnetic field slightly influences the shape of a field line separated by an angle of nearly 90°. Though we may expect that electric currents are distributed within the majority of the expanding CME volume. So, some moving source of the magnetic field in the corona is at an angular distance of only 20° - 30° away from the streamer. That is why its deformation propagates synchronously with the neighbor flank of the CME. Besides, the flux rope is not detached from the Sun but keeps two legs of the expanding loop anchored in the photosphere. The magnetic field of the legs persists in the low corona for a long time, although gradually fading. Its influence on the streamer also decreases.

## 4. Conclusions

We have analyzed the event of the prominent streamer deflection by a CME on 2 January 2012. We have measured the speed of the kink disturbance propagation along the streamer and compared it with the speed of the CME. The kink slightly accelerates in the field-of-view of the LASCO C2 coronagraph and slowly decelerates at greater distances. The leading edge of the CME, moving in SWW direction, accelerates from 840 km s$^{-1}$ to 1100 km s$^{-1}$ and then shows nearly a constant speed with a small deceleration. The northern flank of the CME moves practically synchronously with the streamer kink. We did not find any evidence of streamer axis oscillations. The streamer axis deviates impulsively and then goes back very gradually to the initial position. This behavior does not support the idea of streamer waves as well as the impact on the streamer by a propagating shock wave.

The observed streamer deflection can be interpreted as the influence of the magnetic field of a moving flux rope associated with a CME. The motion of the large-scale flux rope away from the

Sun creates changes in the structure of surrounding field lines, which are similar to the kink propagation along coronal rays.

**Acknowledgements**

This work was supported in part by the Russian Foundation for Basic Research (grants 12-02-00008 and 14-02-92690), by the Program # 22 of the Russian Academy of Sciences, and by the Department of Science and Technology, Ministry of Science and Technology of India (INT/RFBR/P-38, INT/RFBR/P-165).